\providecommand{\e}[1]{\ensuremath{\times 10^{#1}}}

\documentclass[preprint]{aastex}

\begin{document}

\title{A relationship between supermassive black hole mass and the total gravitational mass of the host galaxy}

\author{Kaushala Bandara\altaffilmark{1,2}, David Crampton\altaffilmark{1,2} and Luc Simard\altaffilmark{1,2}}

\affil{Herzberg Institute of Astrophysics, National Research Council of Canada, 5071 West Saanich Road, Victoria, B.C., V9E 2E7, Canada }
\affil{Department of Physics \& Astronomy, University of Victoria, Elliott Building, 3800 Finnerty Road, Victoria, B.C., V8P 5C2, Canada}

\begin{abstract}
We investigate the correlation between the mass of a central supermassive black hole and the total gravitational mass of the host galaxy ($\rm{M_{tot}}$). The results are based on 43 galaxy-scale strong gravitational lenses from the Sloan Lens ACS (SLACS) Survey whose black hole masses were estimated through two scaling relations: the relation between black hole mass and S$\rm{\acute{e}}$rsic index ($\rm{M_{bh}} - \it{n}$) and the relation between black hole mass and stellar velocity dispersion ($\rm{M_{bh} - \sigma_{\star}}$). We use the enclosed mass within $\rm{R_{200}}$, the radius within which the density profile of the early type galaxy exceeds the critical density of the Universe by a factor of 200,  determined by gravitational lens models fitted to HST imaging data, as a tracer of the total gravitational mass. The best fit correlation, where $\rm{M_{bh}}$ is determined from $\rm{M_{bh} -\sigma_{\star}}$ relation, is $\rm{log\:(M_{bh}) = (8.18 \pm 0.11) + (1.55 \pm 0.31) (log\:(M_{tot}) - 13.0)}$ over 2 orders of magnitude in $\rm{M_{bh}}$. From a variety of tests, we find that we cannot reliably infer a connection between $\rm{M_{bh}}$ and $\rm{M_{tot}}$ from the $\rm{M_{bh}} - \it{n}$ relation. The $\rm{M_{bh} - M_{tot}}$ relation provides some of the first, direct observational evidence to test the prediction that supermassive black hole properties are determined by the halo properties of the host galaxy. 
\end{abstract}

\keywords{black hole physics, gravitational lensing, galaxies: evolution, galaxies: halos, quasars: general}

\section{INTRODUCTION}

Supermassive black holes (SMBH) are believed to reside in nearly all galaxies \citep{kor95,fer05} and the masses of these SMBH ($\rm{M_{bh}}$) show correlations with host galaxy properties, implying that SMBH and galaxy formation processes are closely linked \citep{ada01,cat01,cat99,dim05, dim03, el03,hae00,hop05a, hop05b, sil98,wyi06}. Previous studies have shown correlations between $\rm{M_{bh}}$ and galaxy's effective stellar velocity dispersion ($\rm{\sigma_{\star}}$), bulge luminosity ($\rm{L_{bul}}$), S$\rm{\acute{e}}$rsic index ({\it n}) and stellar mass of the bulge component ($\rm{M_{bul}}$) \citep{fer00, geb00, gra01, gra07,mar03}. Some of the challenges faced by current models of SMBH formation and evolution include reproducing and maintaining these scaling relations regardless of the events that take place during galaxy evolution driven by the process of hierarchical mass assembly \citep{cro09,wyi02,wyi03,mcl06,rob06}. These scaling relations are not only important tests of the models of SMBH formation and evolution but also powerful predictive tools from which we can infer SMBH masses of galaxies that are located at higher redshifts.

In this paper, we examine the evidence for a scaling relation between $\rm{M_{bh}}$ and the {\it total} mass of the host galaxy (denoted as $\rm{M_{tot}}$). Most self regulating theoretical models of SMBH formation predict a fundamental connection between $\rm{M_{bh}}$ and $\rm{M_{tot}}$ of the host galaxy \citep{ada01,el03,hae98,mon00,sil98}. One of the most important predictions of the galaxy models, which study the interaction between the dark matter haloes of galaxies and baryonic matter settling into the gravitational potential to form the bulge and SMBH, is that halo properties determine those of the bulge component and SMBH \citep{cat01,el03,hop05a,hop05b}. However, observational evidence for such a scaling relation has been sparse since measurement of the total mass is non-trivial. 

Results from \citet{fer02} examined some of the first {\it indirect} observational evidence for the existence of a $\rm{M_{bh} - M_{tot}}$ relation. In \citet{fer02}, the correlation between bulge velocity dispersion ($\rm{\sigma_{c}}$) and the observed circular velocity ($\rm{v_{c,obs}}$), for a sample of 20 elliptical galaxies and 16 spiral galaxies, is translated into an equivalent $\rm{M_{bh} - M_{tot}}$ correlation. Although $\rm{\sigma_{c}}$ can be translated into $\rm{M_{bh}}$ through $\rm{M_{bh}\:-\sigma_{\star}}$ relation in a straightforwards manner, an estimate of $\rm{M_{tot}}$ is dependent on the conversion between $\rm{v_{c,obs}}$ and virial velocity, the velocity of the galactic halo at the virial radius. \citet{fer02} uses results from $\rm{\Lambda}$CDM cosmological simulations to derive $\rm{M_{tot}}$. Following \citet{fer02}, several studies have examined the $\rm{v_{c,obs} - \sigma_{c}}$ relation and its implications for galaxy formation and evolution \citep{bae03,buy06,cou07,piz05}.

The primary goal of this study is to extend the observational evidence to support the $\rm{M_{bh}\:-M_{tot}}$ relation using an attractive, alternative, and {\it direct} method to measure the total mass of the host galaxy. In recent years, strong gravitational lensing has emerged as a powerful tool to probe the mass profiles, ranging from individual galaxies to clusters of galaxies \citep{bol06,bol08,dye07,gav08,hal08,mou07}. Due to the fact deflection of a photon passing an intervening massive object is independent of the deflector's dynamical state, gravitational lensing does not suffer from difficulties associated with dynamical mass measurements of galaxies, where assumptions regarding orbital motions of tracers can lead to complications. Thus, gravitational lensing is a more robust method to estimate the total mass, including dark matter, around early-type galaxies and infer the existence of an isothermal mass profile ($\rho \propto {r}^{-2}$) in various systems \citep{dye07,koo06}

The mass enclosed within the Einstein radius ($\rm{M_{einst}}$) which is measured from the lens model that produces the best fit to the observed multiple images, is a direct probe of both luminous and dark mass in a galaxy. In conjunction with velocity dispersion and surface brightness profiles of the lens galaxy, measurement of $\rm{M_{einst}}$ from a lens model can be used effectively to constrain the luminous and dark matter profiles. In this study, we utilize the $\rm{M_{einst}}$ as a tracer of the total mass and use the ${\it total}$ mass profile of the lensing galaxy to determine the mass contained within a redshift-independent circular aperture.

We also attempt to derive the $\rm{M_{bh} - M_{tot}}$ relation using purely photometric tracers of $\rm{M_{bh}}$ and $\rm{M_{tot}}$. Therefore, we examine the possibility of using $\rm{M_{bh}} - \it{n}$ relation \citep{gra01, gra07} as the primary method to estimate black hole masses. Surface brightness profiles of the bulge component of most galaxies, in particular E/S0, can be described by the S$\rm{\acute{e}}$rsic law \citep{ser68} as follows:
\begin{eqnarray}
\rm{\Sigma(r) = \Sigma_{e}\exp{(-k[(r/r_{e})^{1/n} - 1]})}
\label{eq:sersic}
\end{eqnarray}
where $\rm{\Sigma}$(r) is the surface brightness at radius $\it{r}$ and S$\rm{\acute{e}}$rsic index (${\it n}$) is the degree of light concentration. Previous studies in the literature indicate that the quantity {\it n} varies monotonically with galaxy magnitude. Therefore, the existence of a $\rm{M_{bh}} - {\it n}$ relation can be inferred given the dependence of $\rm{M_{bh}}$ on the galaxy magnitude \citep{mar03} and the connection between galaxy magnitude and S$\rm{\acute{e}}$rsic index \citep{jer00,gra03,fer06}. 

\cite{gra07} discuss the most recent version of the $\rm{M_{bh}} - {\it n}$ relation and indicate that the dependence between the quantities $\rm{\log(M_{bh})}$ and $\log({\it n})$ can be best represented by a log-quadratic relation. The main motivation for using this relation is that measurement of {\it n} requires only imaging data, which is easier to acquire than spectroscopic data at a given redshift; therefore, $\rm{M_{bh} - M_{tot}}$ could be easily extended to large samples of higher redshift gravitational lenses, which may not have spectroscopic measurements. Furthermore, {\it n} is a distant-independent quantity, an added advantage for estimating black hole masses of high redshift galaxies that may not have secure redshift measurements.

This paper is structured as follows. In \S 2 we present a brief description of the target selection. In \S 3 we describe the following analysis procedures: \S 3.1 Measurements of $\rm{M_{tot}}$; \S 3.2: Deriving the connection between $\rm{M_{bh}}$ and $\rm{M_{tot}}$ using $\rm{M_{bh} - \sigma_{\star}}$ relation; \S 3.3: Deriving the connection between $\rm{M_{bh}}$ and $\rm{M_{tot}}$ using $\rm{M_{bh}} - \it{n}$ relation. In \S 4 we present the results of our analysis. In \S 5 and \S 6 we discuss the implications of our results and give some concluding remarks regarding this project. We assume the following cosmological terms for all computations in the paper:  $\rm{\Omega_{M} = 0.3}$, $\rm{\Omega_{\Lambda} = 0.7}$, $\rm{H_{0} = 70\:h_{70}\:km\:s^{-1}\:Mpc^{-1}}$ and $\rm{h_{70} = 1}$. Unless otherwise noted all scaling relations in this paper are defined as linear relationships in log-log space and all logarithms assume a base of 10.

\section{THE SAMPLE}

Our sample is a subset of the galaxy-scale strong gravitational lenses from the Sloan Lens ACS (SLACS) Survey \citep{bol06,bol08}, a Hubble Space Telescope (HST) imaging survey that is optimized to detect bright early-type lens galaxies with faint lensed sources. These lenses were initially selected from spectra of galaxies of the SDSS Luminous Red Galaxy (LRG) and MAIN galaxy samples \citep{eis01,str02}. After a best-fit template of the continuum was subtracted from the spectrum of each galaxy, the residual spectra were scanned for nebular emission lines at a redshift higher than that of the target galaxy. Galaxies whose residual spectra exhibited at least three atomic transitions at a common background redshift were identified as lens candidates \citep{bol06,bol08} and imaged using the Wide-Field Channel (WFC) of the Advanced Camera for Surveys (ACS) through F435W and F814W filters (Programs 10174, 10587 \& 10886). A detailed description of the gravitational lenses discovered by the SLACS Survey and subsequent analysis can be found in the series, The Sloan Lens ACS Survey 1-8 \citep{bol06, bol08, bol08b, gav07, gav08, koo06, tre06, tre09}. To date, SLACS survey has discovered 131 galaxy-scale strong gravitational lenses with snapshot or deep HST ACS imaging, secure spectroscopic redshifts for both lens and source galaxies in each system and stellar velocity dispersion, $\rm{\sigma_{SDSS}}$, measured within the 3" SDSS spectroscopic fiber \citep{bol08}. 

From the 131 SLACS lenses, 70 systems are classified as ``Grade-A'' systems by \citet {bol08}, where the Grade-A classification implies a definite lens as determined by HST imaging. We use 43 ``Grade-A'' lenses, from programs 10174, 10587 and 10886, as the sample for our analysis. We do not use lens systems that do not have a simple lens model due to nearby companion galaxies (7 lens systems) or systems where $\sigma_{\rm SDSS}$ value is not available (14 lens systems) because the median signal-to-noise (SNR) was too low within the 3" SDSS spectroscopic fiber \citep{bol08}.

\section{DATA ANALYSIS}

The most important advantage of using the SLACS galaxy sample to determine the $\rm{M_{bh} - M_{tot}}$ relation is that we have an accurate measurement of the total mass profile of each galaxy. However, SLACS lenses are located at distances ($\rm{\langle z_{lens}\rangle \approx 0.2}$) where the SMBH mass cannot be measured directly through its influence on the surrounding gas or stars. Therefore, we use two scaling relations to determine the SMBH masses of the galaxy sample: $\rm{M_{bh} - \sigma_{\star}}$ relation \citep{fer00, geb00,gul09} and $\rm{M_{bh}} - \it{n}$ relation \citep{gra01, gra07}.

\subsection{Deriving Total Mass of the Host Galaxy}

We use parameters from the strong lens modeling performed by \citet{bol08} to obtain $\rm{M_{tot}}$. \citet{bol08} use a singular isothermal ellipsoid (SIE) lens model to describe foreground lens plane which generates the multiple images of the source. The SIE lens model is parameterized by the angular Einstein radius, $\rm{b_{SIE}}$, which relates to the physical parameters of the mass model as follows:
\begin{eqnarray}
\rm{b_{SIE} = 4\:\pi\:\frac{\sigma_{SIE}^{2}}{c^{2}}\:\frac{D_{LS}}{D_{OS}}}
\label{eq:SIEprofile}
\end{eqnarray}
where $\rm{D_{LS}}$ and $\rm{D_{OS}}$ are angular diameter distances between the lens and source plane and observer and source plane respectively. The surface brightness profile of the background galaxy is described by a single or multiple S$\rm{\acute{e}}$rsic ellipsoid profile and the model lensed image, for a given SIE lens model, is produced by the forward ray tracing method. The model parameters are adjusted until the images match the observations and the solutions are optimized using a merit function, such as $\rm{\chi^{2}}$ \citep{bol08}.

We use the lens model parameters given by \citet{bol08}, i.e. $\rm{b_{SIE}}$, and integrate the best-fit mass profile to obtain the total mass (both luminous and dark) within a projected radii of choice. During the preliminary analysis, we use the inverted lensing equation to compute mass inside the Einstein radius ($\rm{M_{einst}}$), as a tracer of $\rm{M_{tot}}$, assuming that the lens and source galaxies are aligned along the optical axis. 

However, $\rm{M_{einst}}$ is not a measurement of mass within a ${\it standard}$ aperture since the projected radius characterized by the Einstein radius for a given mass distribution is redshift dependent. We therefore use $\rm{R_{200}}$, the radius within which the mean density of the lensing galaxy exceeds the critical density of the Universe ($\rho_{\rm crit}$) by a factor of 200, as a ``standard aperture''. The use of $\rm{R_{200}}$ as a radial aperture is also consistent with theoretical studies of the $\rm{M_{bh} - M_{tot}}$ relation \citep{cro09}. We integrate the best-fit mass profile to obtain the mass contained within a projected radius of  $\rm{R_{200}}$ ($\rm{M_{200}}$) as a tracer of the total mass as characterized below:
\begin{eqnarray}
\rm{M_{200} \approx \frac{2\:(\sigma_{SIE})^{2}\:R_{200}}{G}}
\label{eq:totalmass}
\end{eqnarray}

Adopting an ``aperture-corrected" mass measurement ($\rm{M_{200}}$), instead of $\rm{M_{einst}}$, does not affect the overall form of the $\rm{M_{bh} - M_{tot}}$ relation (the slope and zero-point) discussed in the following sections. Results of \citet{bol08b} show a lack of correlation between the quantities $\rm{f \equiv \sigma_{aperture} / \sigma_{SIE}}$ and mass or $\rm{R_{einst} / R_{aperture}}$, where $\rm{\sigma_{aperture}}$ is the stellar velocity dispersion corrected to an aperture of $\rm{R_{aperture}}$. These results are consistent with the near isothermal nature of the radial profile; thus, using a radial aperture other than $\rm{R_{einst}}$, to derive the total mass, does not introduce an additional dependence on the assumed mass model. We also observe a relationship of $\rm{log\:(M_{einst} / M_{200}) \propto log\:(R_{einst} / R_{200})}$, consistent with the fact that an aperture corrected mass measurement does not affect the observed $\rm{M_{bh} - M_{tot}}$ relation. 

To obtain the associated 1-$\sigma$ error of $\rm{M_{200}}$, we use Gaussian error propagation adopting an empirical error of 2\% on the measured Einstein radii following \citet{bol08} ($\langle\delta\log(\rm{M_{200}})\rangle \approx 0.01\:\rm{dex}$). Table 1 lists the unique SDSS identifiers for SLACS lenses, redshift of the lens galaxy of a system and the derived values of $\rm{M_{200}}$ (hereafter denoted as $\rm{M_{tot}}$) for each lensing system.

\subsection{Estimating SMBH Masses Using Stellar Velocity Dispersion}

We first use the $\rm{M_{bh} - \sigma_{\star}}$ relation to estimate $\rm{M_{bh}}$ of the early type lens sample. The velocity dispersion measurements are derived from spectra from the SDSS 3" diameter fiber. The SDSS velocity dispersion measurements are corrected to a standard aperture, of radius equal to $r_{\rm e}\:/\:8$, using the power-law relation by \citet{jor95}, similar to SMBH and Fundamental Plane studies \citep{fer05, tre06}. The mean difference between SDSS velocity dispersion measurements and the aperture corrected measurements (hereafter denoted as $\rm{\sigma_{\star}}$) are $~\approx4\%$; therefore, the aperture correction does not significantly influence the overall form of the $\rm{M_{bh} - M_{tot}}$ relation discussed in the following sections. We use $\rm{M_{bh} - \sigma_{\star}}$ relation by \citet{gul09}, obtained from measurements of $\rm{M_{bh}}$ from dynamically detected central black holes:
\begin{eqnarray}
\rm{log(M_{bh} / M_{\odot}) = (8.12 \pm 0.08) + (4.24 \pm 0.41)\:log(\sigma_{\star} / 200\:km\:s^{-1})}
\label{eq:Msigma}
\end{eqnarray}
with an intrinsic scatter of $\rm{\varepsilon_{0}} = 0.44$ dex.

We assume that errors on $\sigma_{\star}$ and best-fit coefficients of the $\rm{M_{bh}\:-\:\sigma_{\star}}$ relation are uncorrelated and use Gaussian error propagation to determine the 1-$\sigma$ error on the quantity $\log(\rm{M_{bh}})$. The values of $\sigma_{\star}$ and its associated error for each lensing system and SMBH mass estimates obtained from $\rm{M_{bh} - \sigma_{\star}}$ relation are listed in Table 1.

\subsection{Estimating SMBH Masses Using the S$\rm{\acute{e}}$rsic Index}

We also attempt to use the $\rm{M_{bh}} - {\it n}$ relation to estimate $\rm{M_{bh}}$ of the SLACS lens sample \citep{gra01,gra07}. We perform a careful and detailed decomposition of HST ACS images to obtain the S$\rm{\acute{e}}$rsic indices of the SLACS lens galaxies. Although our sample of lens galaxies were observed in at least two ACS filters, mainly F814W and F435W, we exclusively use F814W (Broad I-band) data due to high SNR of the images. Furthermore, redder magnitudes are a better tracer of mass in comparison to B-band magnitudes. To keep our decomposition method consistent with that of \citet{gra07}, we use a two-component model to describe the surface brightness profiles of SLACS lens galaxies. In addition to the S$\rm{\acute{e}}$rsic profile, which describes the surface brightness of the bulge component in each galaxy, we include a second disk component characterized by a simple exponential profile:
\begin{eqnarray}
\rm{\Sigma(r) = \Sigma_{0}\exp(-r/r_{d})}
\label{eq:disk}
\end{eqnarray}
where $\Sigma(r)$ is the surface brightness at radius $\it{r}$. 

To obtain the best possible accuracy of the data analysis process, we take particular care during all intermediate steps leading to the bulge-disk composition. Therefore, we obtain the raw images and perform careful cosmic ray removal, distortion correction, manual mask production and determination of the best possible ACS point-spread function (PSF). We obtain the ACS images from the HST archive at The Canadian Astronomy Data Centre (CADC), where the images are processed by CALACS calibration software. Imaging data from programs 10174 and 10587 are 420 s single-exposure snapshot data; therefore, we perform an additional cosmic ray identification step using the L.A.Cosmic software (LACOS) \citep{van01}, which distinguishes between undersampled sources and cosmic rays. We then use the MULTIDRIZZLE reduction package to obtain distortion-free ACS images, where the distortion solution is applied to correct all pixels to equal areas.

We use the photometry package SExtractor \citep{ber96} to produce a catalog of galaxies in each ACS image. Furthermore, we use the segmentation images generated by SExtractor, which deblends each source in the field, to produce manual masks of the lensed features around the early type galaxy before lens modeling. We perform the two-dimensional decomposition of each early-type lens galaxy using Galaxy IMage 2D (GIM2D), which is publicly available to users \citep{sim02}. GIM2D uses the Metropolis Algorithm to derive the best-fit parameter values and confidence intervals, for a surface brightness model of choice, through Monte-Carlo sampling of the likelihood function. Since our surface brightness model consist of a bulge and a disk component, we explore the full range of bulge fraction (B/T = 0.0 - 1.0) such that the Metropolis Algorithm converges to an accurate quantitative morphology of galaxies classified as E-types. On average, we find that most SLACS lenses have a significant bulge component ($\rm{\langle\:B/T\:\rangle \approx 0.73}$) and a bulge plus disk light profile produces a better fit to the lens galaxy, significantly minimizing the residuals when the galaxy model is subtracted from the input galaxy image. 

\begin{figure}
\includegraphics[width=\textwidth]{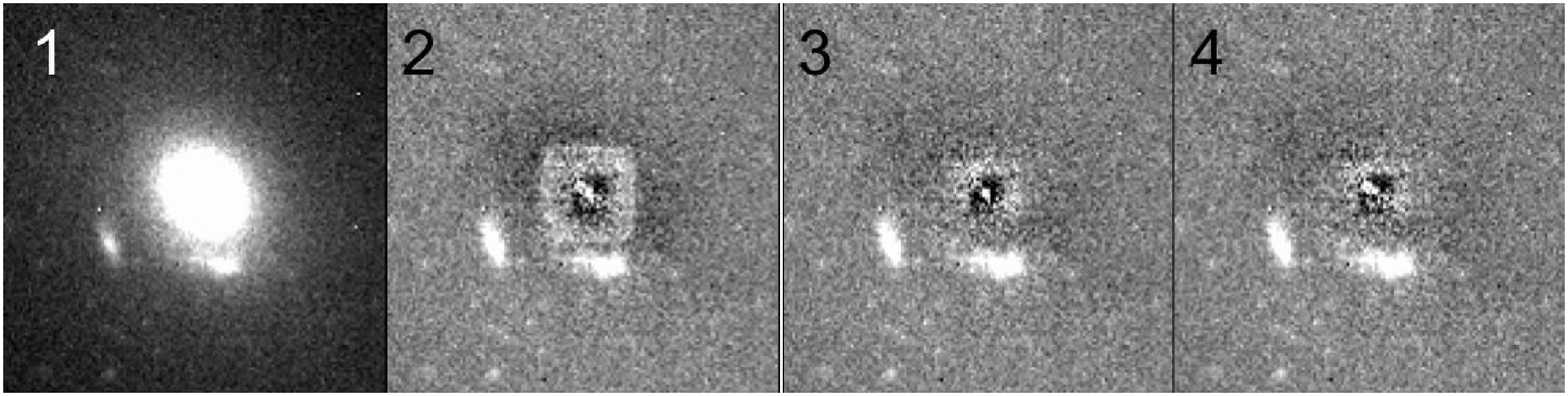}
\caption{Comparison of varying PSFs on GIM2D galaxy models. Panel 1 shows the HST ACS image of an example SLACS lens. The contrast of panel 1 is exaggerated to show the lensing features around the early-type galaxy. Panels 2, 3, and 4 show the residual images after GIM2D surface brightness model (S$\rm{\acute{e}}$rsic bulge + exponential disk) convolved with a multidrizzled Tiny-Tim PSF (panel 2), natural PSF (panel 3) and ACS-library generated PSF (panel 4) is subtracted from the observed image. Residual image where the galaxy model is convolved with a Tiny-Tim PSF shows artifacts such as boxy-core feature, due to the finite size of the Tiny-Tim PSF, evident in panel 2. Statistics of the pixels in panel 2 show that the mean value of the core residuals is comparable to the background level; therefore, the artifact introduced by the finite Tiny-Tim PSF does not significantly affect the quality of the overall fit. In general, we find that both natural and ACS-library generated PSFs produce minimal core residuals. However, most SLACS fields are relatively devoid of stars, suitable for convolution; therefore, we use ACS-library generated PSF for GIM2D galaxy modeling.}
\label{fig:psfcompare}
\end{figure}

For GIM2D lens models, we take particular care to define the ACS Point Spread Function (PSF), which is extremely position- and time- dependent. We investigate a variety of PSF models to determine the most suitable PSF for convolution with GIM2D galaxy models. The PSF models are as follows.
\begin{enumerate}
\item A star extracted from the field.
\item A PSF produced at the location of the galaxy, in the non-drizzled image, using Tiny Tim software \citep{kri93}. We insert the Tiny Tim generated PSF into an empty ACS-WFC field, at the location of the galaxy, and multidrizzle the resulting image to obtain a distortion-free PSF model. 
\item A PSF produced at the location of the galaxy using the principle component analysis (PCA) models in the ACS PSF library \citep{jee07}.
\end{enumerate}
We examined the residual images produced by subtracting the lens surface brightness model, each convolved with the PSF models described above, to determine the best-fit PSF for GIM2D lens modeling. The PSF model produced from the ACS library gives the best results, minimizing the core residuals for a fixed surface brightness model. Therefore, we extract the S$\rm{\acute{e}}$rsic indices from the best-fit bulge plus disk decomposition, using the ACS PSF library model for convolution. An example of the difference between residual images produced by various PSF models are shown in Figure~\ref{fig:psfcompare}. 

\begin{figure}
\includegraphics[angle=270, width=\textwidth]{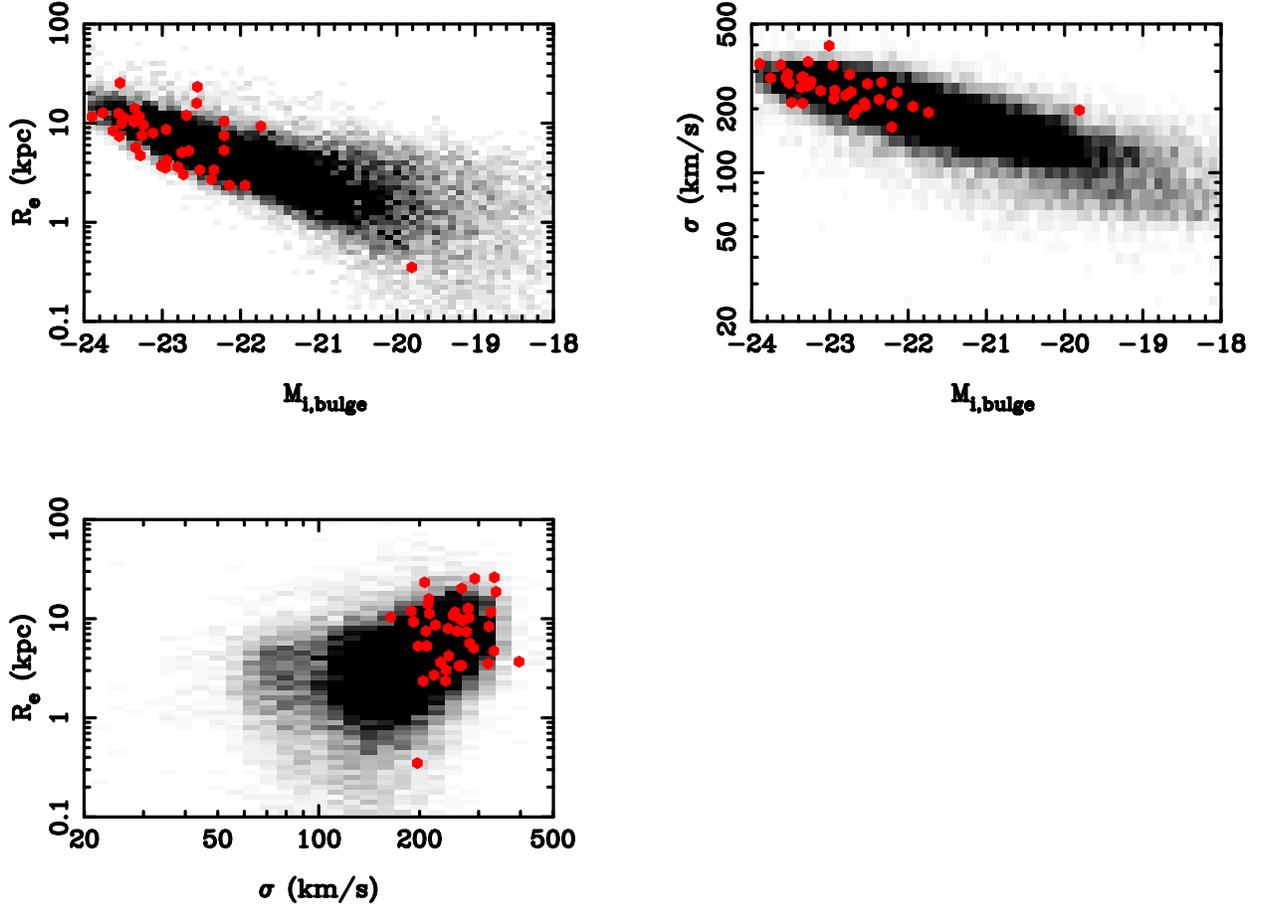}
\caption{Comparison of SLACS lenses to the scaling relations for bulge components of SDSS galaxies. The red circles indicate the best-fit values for the following structural parameters of the SLACS lenses: effective bulge radius ($\rm{R_{e}}$), I-band magnitude of the bulge ($\rm{M_{I,bulge}}$) and aperture corrected bulge velocity dispersion ($\rm{\sigma}$). The scaling relations between bulge structural parameters for SDSS early type galaxies are constructed from a sample 77523 galaxies and are shown in gray-scale. The gray-scale is scaled to 20\% of the peak value of the central distribution to show that the structural parameters of the SLACS lenses lie well within the distribution that encompasses the SDSS galaxies. The outlier in this figure is SDSS J0959 + 0410.}
\label{fig:kormendy}
\end{figure}

We also examine the reliability of the bulge-disk decomposition by comparing our results for the SLACS lens sample relative to the scaling relations for spheroidal components of the SDSS galaxies. Since SLACS lenses were initially derived from SDSS Luminous Red Galaxy (LRG) and MAIN samples, structural parameters measured from ACS imaging should be in agreement with those of SDSS galaxies. Figure~\ref{fig:kormendy} shows the comparison of effective bulge radius ($\rm{R_{e}}$), I-band magnitude of the bulge component ($\rm{M_{I,bulge}}$) and aperture-corrected bulge velocity dispersion ($\rm{\sigma_{ap}}$) of the SLACS lenses to the SDSS bulge parameters. We construct the scaling relations for SDSS galaxies from the bulge-disk decomposition of 77523 galaxies and velocity dispersions from the SDSS database. The structural parameters derived from ACS imaging are in good agreement with those of SDSS galaxies, which confirms the accuracy of our bulge-disk decomposition (in addition to the minimal residuals seen in the galaxy images after subtracting the galaxy models). The outlier in Figure \ref{fig:kormendy} is SDSS J0959 + 0410, a disk galaxy with a small bulge fraction ($\rm{B/T = 0.14}$). Previous studies, which examine the positions of bulges with disk-like features (also referred to as $\it{pseudobulges}$) in structural parameter space find that pseudobulges with low $\rm{B/T}$ deviate from the scaling relations of classical bulges \citep{fis08}. 

Using the best-fit parameter for {\it n} from GIM2D models, we estimate the SMBH masses using the log-quadratic $M_{\rm bh} - n$ relation by \citep{gra07}:
\begin{eqnarray}
\rm{log (M_{bh} / M_{\odot})} = (7.98 \pm 0.09) + (3.70 \pm 0.46) \log({\it n}/3) - (3.10 \pm 0.84) [ \log({\it n}/3) ]^2
\label{eq:GrahamMn}
\end{eqnarray}
with an intrinsic scatter of $\rm{\varepsilon_{0}} = 0.18$ dex. 

Similar to \citet{gra07}, we assume a measurement error of 20\% on the values of {\it n} to obtain the 1-$\sigma$ error on $\log(\rm{M_{bh}})$. Table 1 lists the S$\rm{\acute{e}}$rsic indices and bulge fractions of the SLACS lenses from GIM2D modeling, estimates of $\rm{M_{bh}}$ for each lens system.

\section{RESULTS}

\subsection{$\rm{M_{\rm{bh,\sigma_{\star}}}\:-\:M_{\rm{tot}}}$ Relation}

\begin{figure}
\includegraphics[width=\textwidth]{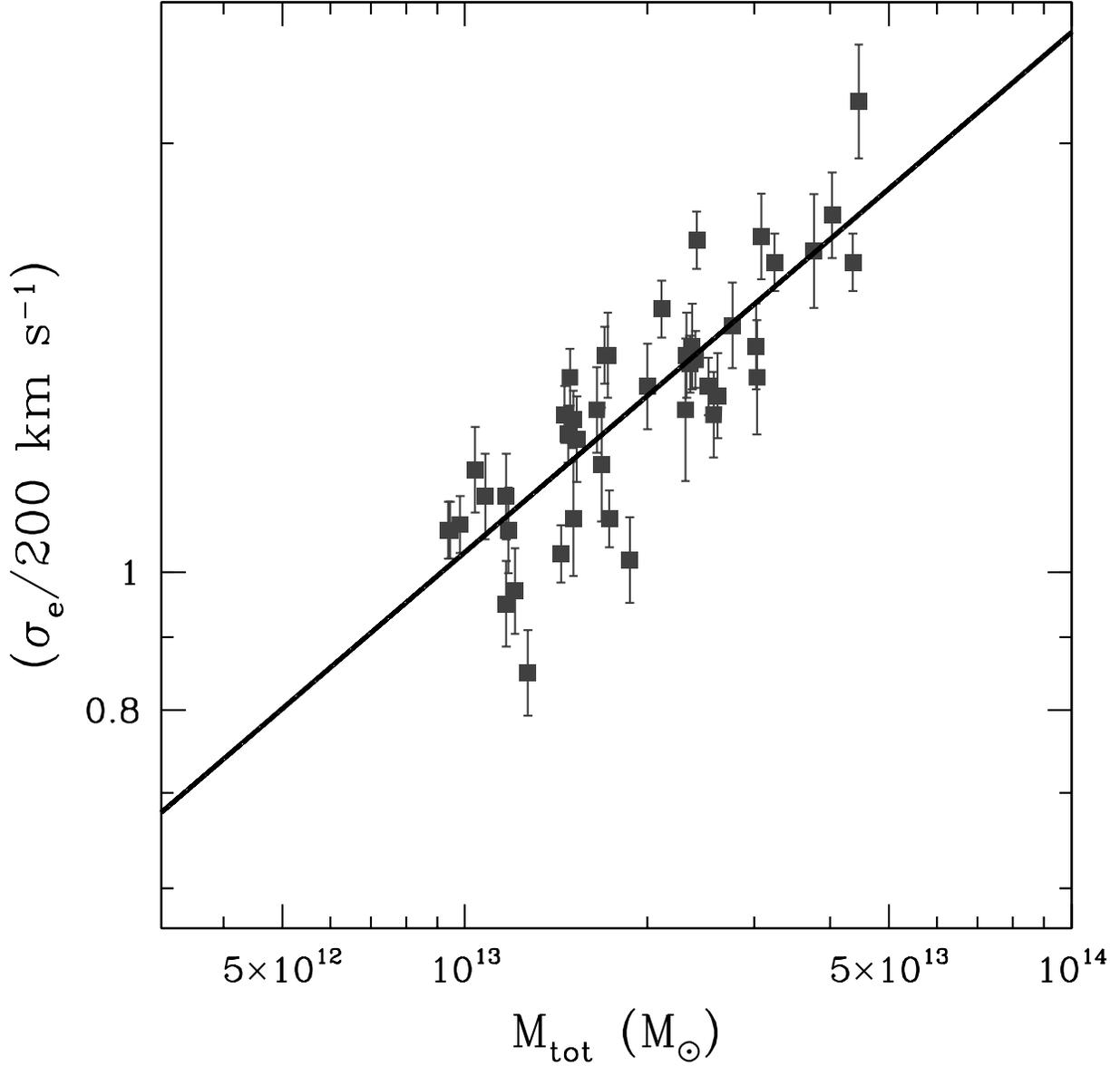}
\caption{The correlation between $\rm{\sigma_{\star}}$ and $\rm{M_{tot}}$, for the sample of early type SLACS lenses. Velocity dispersion values from SDSS pipeline are corrected to an aperture of radius equal to $\rm{r_{e}/8}$ using the empirical relation \citet{jor95}. The error bars correspond to the 1-$\rm{\sigma}$ error of the aperture corrected velocity dispersion, from Gaussian error propagation, taking the measurement errors of SDSS velocity dispersions into account. Total masses of the SLACS lenses are derived from strong lens modeling parameters of \citet{bol08}. The 1-$\rm{\sigma}$ errors of $\rm{M_{tot}}$, which are smaller than the data points, are incorporated into the fitting routines but not shown in this plot. The solid line correspond to the best-fit $\rm{\sigma_{\star} - M_{tot}}$ relation for all SLACS lenses in the logarithmic space.}
\label{fig:sigmamtot}
\end{figure}

In the following section, we combine the correlation between the primary observable quantities ($\rm{\sigma_{\star} - M_{tot}}$) with $\rm{M_{bh} - \sigma_{\star}}$ relation to derive the $\rm{M_{bh} - M_{tot}}$ relation. A tight correlation between the quantities $\rm{log}(\sigma_{\star}/200\:km\:s^{-1})$ and $\rm{log}(M_{\rm tot})$ is apparent from Figure~\ref{fig:sigmamtot}. From Spearman's rank test, we obtain a correlation coefficient ($r_{\rm s}$) of 0.84 (degrees of freedom = 41), which indicates a strong positive correlation between the quantities $\rm{log}(\sigma_{\star}/200\:km\:s^{-1})$ and $\rm{log}(M_{\rm tot})$, with a $99.99\%$ confidence level that the correlation has not occurred by chance. 

To quantify this correlation, we use the $\rm{\chi^{2}}$-fitting routine by \citet{wei06}, which implements a generalized form of the least-squares fitting routine by \citet{pre92}. \citet{wei06} routine accounts for intrinsic scatter ($\varepsilon_{0}$) beyond the observational errors, for a relation of interest, by adding $\varepsilon_{0}$ in quadrature to the error in the dependent variable. Initial fits to the observed relation between $\rm{log}(\sigma_{\star}/ 200\:km\:s^{-1})$ and $\rm{log}(M_{\rm tot})$, incorporating observational errors in both variables, gives a large reduced $\rm{\chi^{2}}$ value ($\rm{{\chi_{red}}^{2} \approx 3.0}$) indicative of intrinsic scatter in the relation. Therefore, we perform the fits by incorporating observational errors in both variables and intrinsic scatter in the $\it{y}$-variable. The value of $\varepsilon_{0}$ is determined by requiring that $\rm{{\chi_{red}}^{2}}$ is unity. The results of the fitting routine gives the following best fit correlation:
\begin{eqnarray}
\rm{log\:(\sigma_{\star}/ 200\:km\:s^{-1}) = (0.014 \pm 0.013) + (0.365 \pm 0.038)[log\:(M_{tot}/M_{\odot}) - 13.0]}
\label{eq:sigmamtoteq}
\end{eqnarray}
with an intrinsic scatter of 0.037 dex in $\rm{log}(\sigma_{\star}/ 200\:km\:s^{-1})$. Combining equations~\ref{eq:Msigma} and ~\ref{eq:sigmamtoteq}, we derive the following $\rm{M_{bh} - M_{tot}}$ relation:
\begin{eqnarray}
\rm{log\:(M_{bh}/M_{\odot})} = (8.18 \pm 0.11) + (1.55 \pm 0.31)\:[\rm{log\:(M_{tot}/M_{\odot})} - 13.0]
\label{eq:observational}
\end{eqnarray}

Figure~\ref{fig:fromMsigma} shows $\rm{\sigma_{\star}}$ transformed into SMBH masses using equation~\ref{eq:Msigma}. We assume that measurement errors of the velocity dispersions and best-fit coefficients of the $\rm{M_{bh} - \sigma_{\star}}$ relation are uncorrelated and use Gaussian error propagation to determine the 1-$\sigma$ error on the quantity $\rm{log}(M_{\rm bh})$. We also perform a direct fitting step to $\rm{M_{tot}}$ and the secondary observable quantity, $\rm{M_{bh}}$, to confirm the validity of the derived $\rm{M_{bh} - M_{tot}}$ relation. We find $\rm{log}(M_{\rm bh}/M_{\odot}) = \alpha + \beta(\rm{log}(M_{\rm tot}/M_{\odot}) - 13.0)$ with $(\alpha, \beta) = (8.17 \pm 0.13, 1.57 \pm 0.39)$ and $\rm{{\chi_{red}}^{2} \approx 0.2}$, which is in agreement with the result shown in equation~\ref{eq:observational}.  

\subsection{$\rm{M_{\rm{bh,n}}\:-\:M_{\rm tot}}$ Relation}

\begin{figure}
\includegraphics[width=\textwidth]{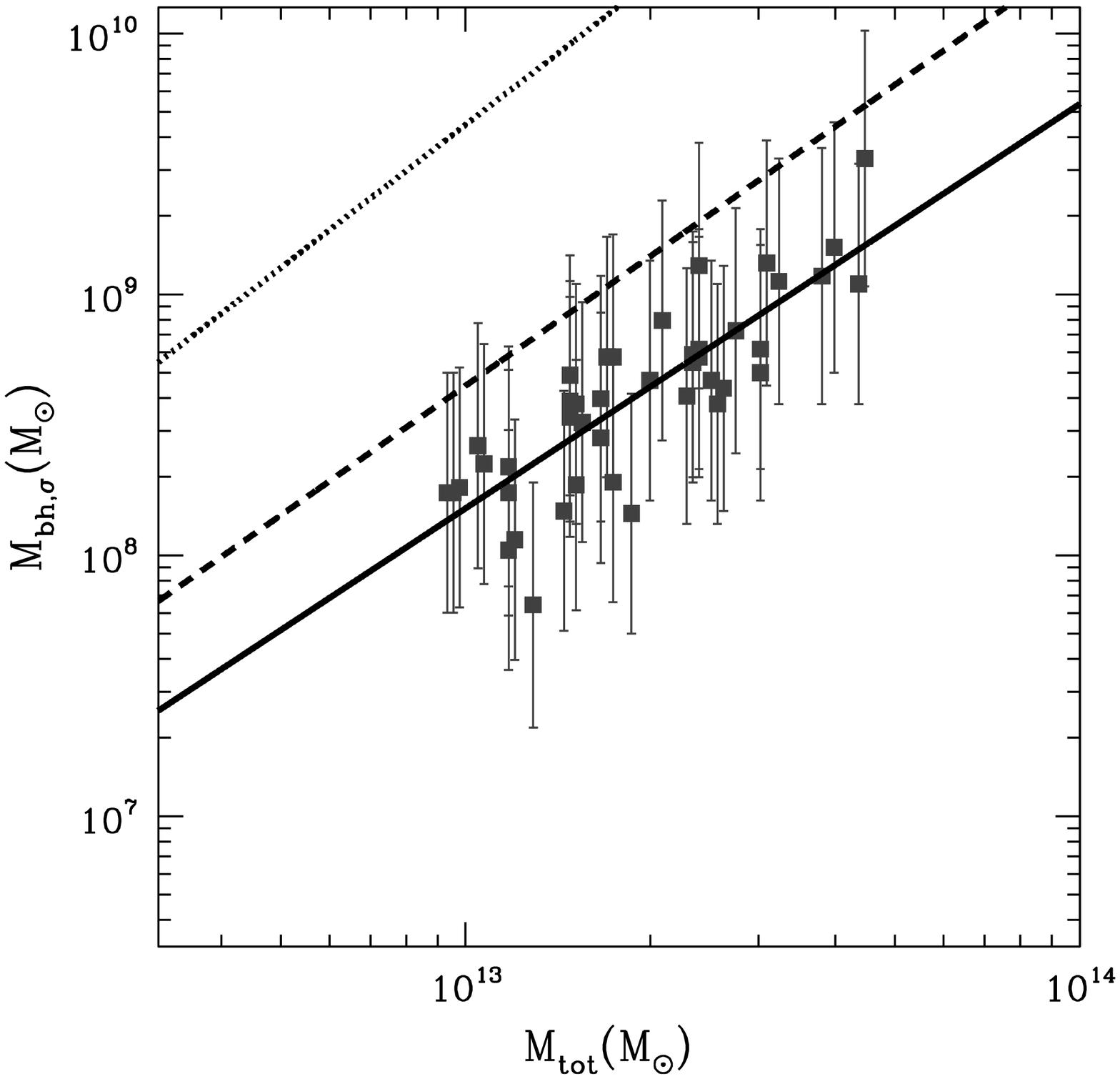}
\caption{Same as Figure~\ref{fig:sigmamtot} with $\rm{\sigma_{\star}}$ converted into $\rm{M_{bh}}$ using $\rm{M_{bh} - M_{tot}}$ relation \citep{gul09}. The error bars correspond to 1-$\sigma$ error on the quantity $\rm{\log(M_{bh}})$. The solid line corresponds to the $\rm{M_{bh} - M_{tot}}$ relation for all SLACS lenses derived using the $\rm{\sigma_{\star} - M_{tot}}$ and $\rm{M_{bh} - M_{tot}}$ relations. The dashed line represents the resulting $\rm{M_{bh}-M_{tot}}$ relation obtained by \citet{fer02}, where $\rm{M_{tot}}$ is computed from cosmological simulations \citep{bul01} relating the the observed circular velocity ($v_{\rm {c,obs}}$) to the virial velocity of the host halo ($v_{\rm{vir}}$). The dotted line shows the resulting $\rm{M_{bh}-M_{tot}}$ relation if $\rm{M_{tot}}$ is computed using $v_{\rm c,obs} = 1.8\:v_{\rm vir}$ \citep{fer02}.}
\label{fig:fromMsigma}
\end{figure}

\begin{figure}
\includegraphics[width=\textwidth]{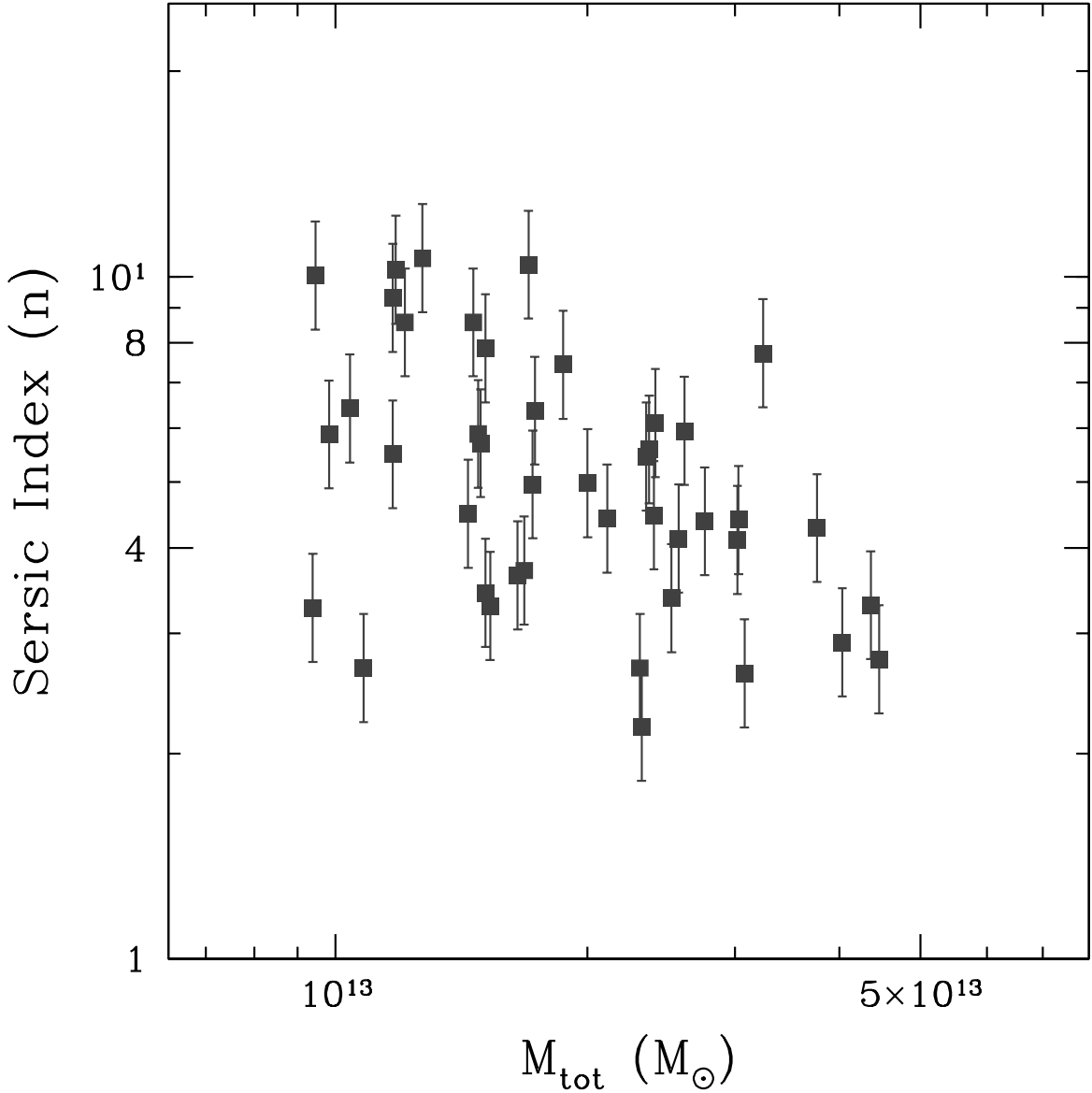}
\caption{The correlation between $n$ and $\rm{M_{tot}}$, for the sample of early type SLACS lenses. The S$\rm{\acute{e}}$rsic indices are derived from the best-fit bulge+disk decomposition of the SLACS lenses. The error bars correspond to the measurement errors on the quantity $n$. Total masses of the SLACS lenses are derived from strong lens modeling parameters of \citet{bol08}. The 1-$\sigma$ errors of $\rm{M_{tot}}$, which are smaller than the data points, are incorporated into the fitting routines but not shown in this plot.}
\label{fig:nmtot}
\end{figure}

\begin{figure}
\includegraphics[width=\textwidth]{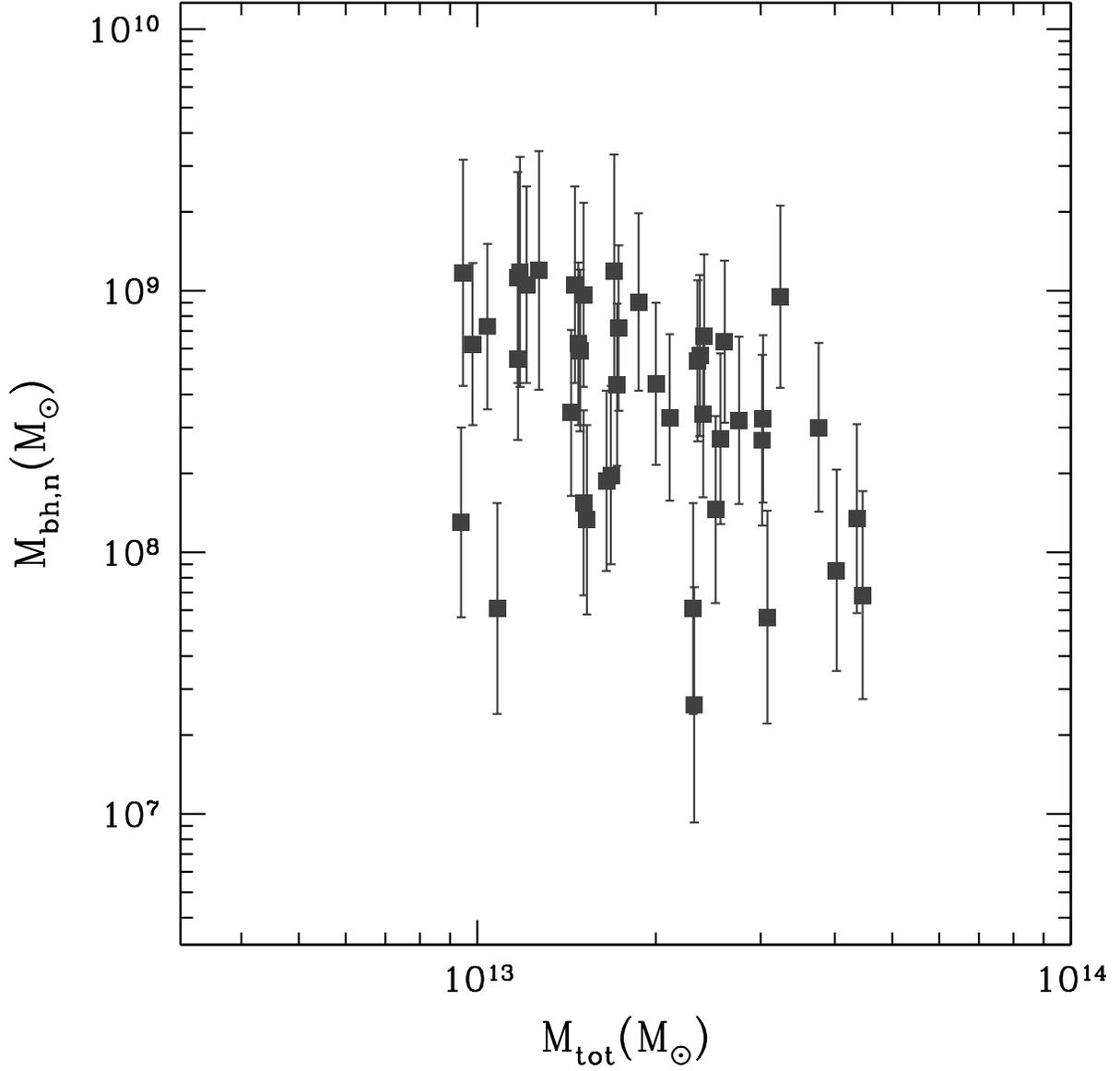}
\caption{Same as Figure~\ref{fig:nmtot} with $n$ converted into $\rm{M_{bh}}$ using $\rm{M_{bh}} - {\it n}$ relation \citep{gra07}. The error bars correspond to 1-$\sigma$ error on the quantity $\rm{\log(M_{bh}})$.The 1-$\sigma$ error of $\rm{M_{tot}}$ is smaller than the data points, and is not shown in the plot.}
\label{fig:fromMn}
\end{figure}

In this section, we discuss the $n - \rm{M_{tot}}$ relation of the SLACS lenses, shown in Figure~\ref{fig:nmtot}, and Figure~\ref{fig:fromMn} which shows $n$ converted to SMBH masses using equation~\ref{eq:GrahamMn} \citep{gra07}. Similar to \S 4.1, we use Gaussian error propagation (assuming a measurement error of 20\% on $n$) with intrinsic scatter of the $\rm{M_{bh} - M_{tot}}$ relation ($\rm{\varepsilon_{0} = 0.18\:dex}$) added in quadrature, to determine the 1-$\sigma$ error of the quantity $\rm{log(M_{bh})}$. Results of  \citet{gra07} indicate that $\rm{M_{bh}} -{\it n}$ relation is comparable to the $\rm{M_{bh} - \sigma_{\star}}$ relation; therefore, we expect both scaling relations to yield $\rm{M_{bh} - M_{tot}}$ relations with similar level of scatter. If the expected equivalence between $\rm{M_{bh}}\:-\:{\it n}$ and $\rm{M_{bh}\:-\:\sigma_{\star}}$ relations hold, we can then extend the $\rm{M_{bh} - M_{tot}}$ relation to gravitational lens samples which span a higher dynamical range in the total mass at various redshifts. Unfortunately, however, the trends observed in Figures~\ref{fig:nmtot} and \ref{fig:fromMn} are very different from the strong positive correlations shown in Figures~\ref{fig:sigmamtot} and \ref{fig:fromMsigma}.

From a Spearman's rank test, we obtain a weak negative $n - \rm{M_{tot}}$ correlation ($r_{\rm s}\approx-0.42$ for degrees of freedom = 41) for the trend observed in Figure~\ref{fig:nmtot} with only a $95\%$ confidence level that the correlation has not occurred by chance. A Spearman's rank test performed on the $\rm{M_{bh} - M_{tot}}$ relation shows similar statistical properties to that of the $n - \rm{M_{tot}}$ relation. Before concluding if the correlation observed in Figure~\ref{fig:fromMsigma} is a better representation of a $\rm{M_{bh} - M_{tot}}$ relation, we investigate various possibilities that may cause the inconsistency observed between Figures~\ref{fig:fromMsigma} and \ref{fig:fromMn}. 

\begin{figure}
\includegraphics[width=\textwidth]{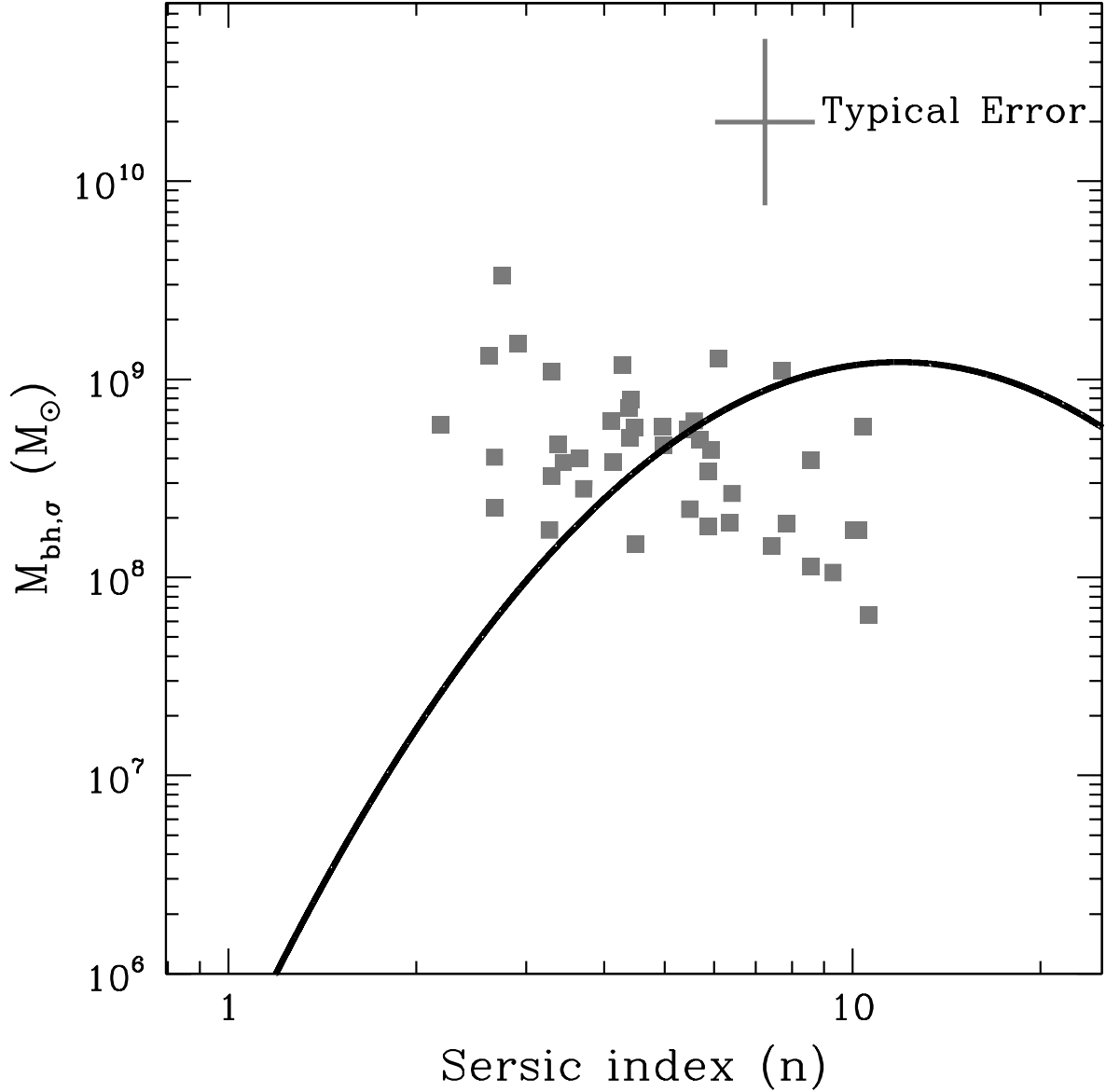}
\caption{Comparison of log-quadratic $\rm{M_{bh}} - {\it n}$ relation to SLACS data. $\rm{M_{bh}}$ values are derived from the $\rm{M_{bh} - \sigma_{\star}}$ relation and S$\rm{\acute{e}}$rsic indices are best-fit results from the GIM2D bulge-disk decomposition. Error bar on the legend represent the typical 1-$\sigma$ uncertainty of $\rm{log\:(M_{bh})}$, derived from Gaussian error propagation using measurement errors from the SDSS velocity dispersions ($\rm{\langle \delta\:log\:(M_{\rm bh}) \rangle \approx 0.42\:dex}$), and a measurement error of 20\% on the quantity ${\it n}$. The log-quadratic relation by \citet{gra07} is shown by the solid black line.}
\label{fig:graham}
\end{figure}

First, we attempted to reproduce the $\rm{M_{bh}} - \it{n}$ relation, obtained by \citet{gra07}, for the SLACS lens sample. In Figure~\ref{fig:graham}, we plot $\rm{M_{bh,\sigma_{\star}}}$ values for the SLACS lenses, versus the corresponding S$\rm{\acute{e}}$rsic indices from the bulge-disk decomposition. The log-quadratic $\rm{M_{bh}} - \it{n}$ relation by \citet{gra07} is overlaid on the SLACS data for comparison. Figure~\ref{fig:graham} clearly indicates that the log-quadratic $\rm{M_{bh}} - n$ relation cannot be reproduced with the $\rm{M_{bh,\sigma_{\star}}}$ and $n$ values for the SLACS lens sample. A weak but different correlation between $\rm{M_{bh,\sigma_{\star}}}$ and ${\it n}$ is apparent from Figure~\ref{fig:graham}. We perform a Spearman's rank test and derive a weak, negative correlation, $r_{\rm s} \approx -0.44$ at the $5\%$ significance level, between these quantities.

The correlation observed in Figure~\ref{fig:graham} is in contrast to the log-linear or log-quadratic $\rm{M_{bh}} - {\it n}$ relation discussed in literature \citep{gra07}. Previous studies of morphology and scaling relations of bulge dominated galaxies claim that ${\it n}$ scales monotonically with the galaxy bulge magnitude (equivalently bulge luminosity) \citep{fer06}. If this correlation holds, we expect ${\it n}$ to correlate positively with $\rm{M_{bh}}$ due the dependence between $\rm{M_{bh}}$ and bulge luminosity ($\rm{L_{bul}}$) \citep{mar03,fer06}. 

\begin{figure}
\includegraphics[width=\textwidth]{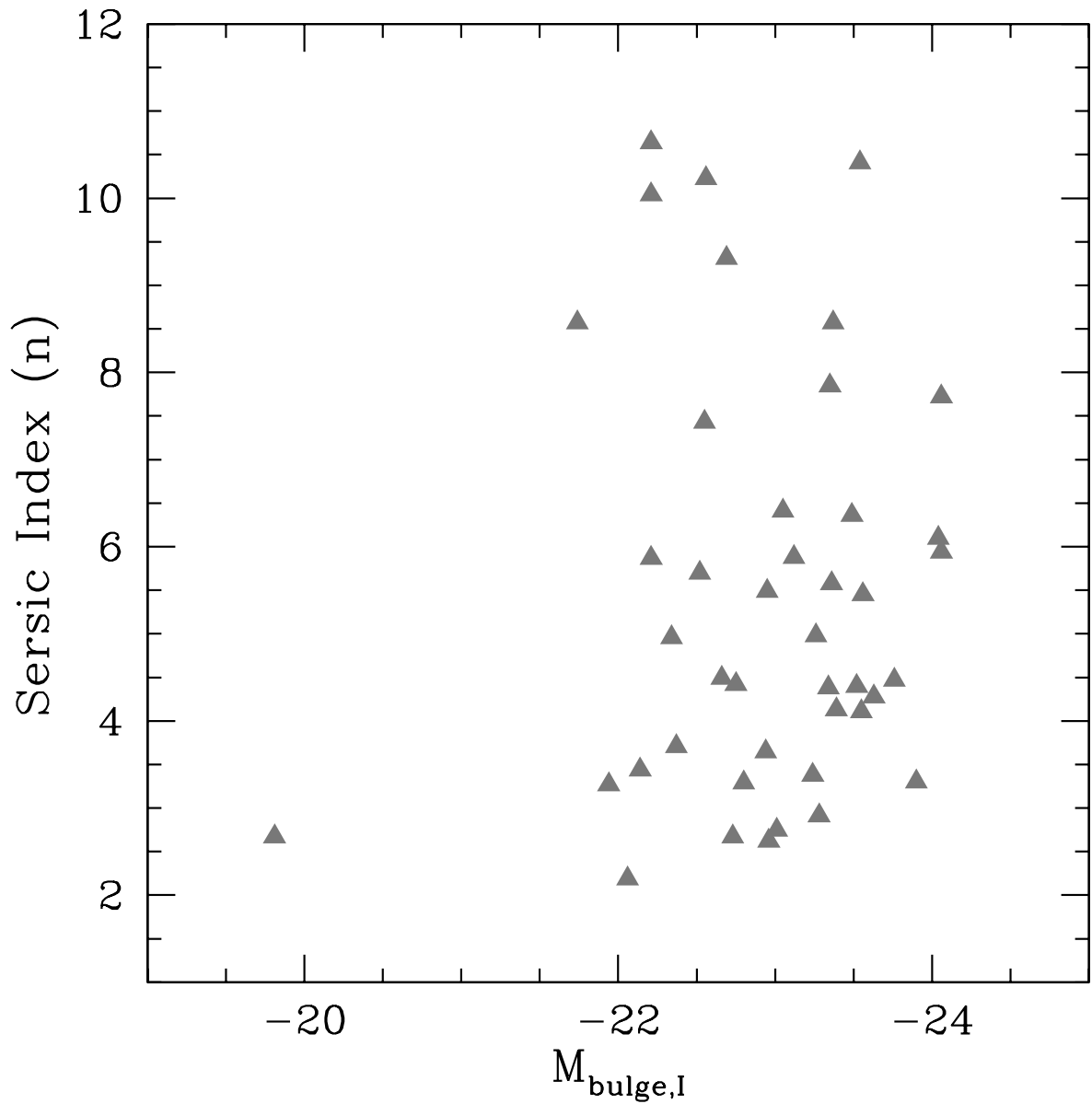}
\caption{The correlation between $\rm{M_{bulge,I}}$ versus $\it{n}$ for SLACS lens sample. The values of $n$ are the best fit $\it{bulge}$ S$\rm{\acute{e}}$rsic indices from the bulge+disk decomposition using GIM2D. The values of $\rm{M_{bulge,I}}$ are the best-fit, extinction-corrected, I-band $\it{bulge}$ magnitudes from the bulge+disk decomposition.}
\label{fig:Magvsn}
\end{figure}

Since Figure~\ref{fig:graham} contradicts the existence of a log-quadratic $\rm{M_{bh}} - {\it n}$ relation, we also test the correlation between ${\it n}$ and I-band bulge magnitude ($\rm{M_{bulge,I}}$) for the SLACS lens sample. The results of GIM2D bulge-disk decomposition of SLACS galaxies are shown in Figure~\ref{fig:Magvsn}. A Spearman's rank test to determine degree of correlation between $\rm{M_{bulge,I}}$ and ${\it n}$ shows that there is only a very weak correlation between the two quantities ($r_{\rm s} \approx -0.20$); furthermore, the rank correlation we observe between $\rm{M_{bulge,I}}$ and ${\it n}$ is significant at a level larger than 5\%.

Although we do not observe a direct correlation between galaxy bulge magnitude and the S$\rm{\acute{e}}$rsic index, our results are limited by the narrow range of magnitudes ($-24\:<\:\rm{M_{bulge,I}}\:<\:-20$) in the SLACS lens sample. Furthermore, our results are consistent with those of \citet{bol08b}, who also find that ${\it n}$ for all SLACS lenses is uncorrelated with the measurements of lensing mass, dynamical mass, luminosity and velocity dispersion of the sample. Given the trend observed in Figure~\ref{fig:Magvsn}, we cannot confidently expect $\rm{M_{bh}}$ to correlate with ${\it n}$ as claimed by previous studies \citep{gra01,gra07}. The possible origin of $\rm{M_{bh}} - {\it n}$ relation has not been examined theoretically and is beyond the scope of this paper. For the remainder of this study, we use the $\rm{M_{bh} - \sigma_{\star}}$ relation to determine the black hole masses. 

\section{TOWARDS A $\rm{M_{\rm BH}\:-\:M_{\rm TOT}}$ RELATION}

The correlation observed between $\rm{M_{bh}}$ and $\rm{M_{tot}}$, in Figure~\ref{fig:fromMsigma}, is in remarkable agreement with theoretical predictions \citep{cro06,cro09,sil98,wyi02,wyi03} and local observations \citep{fer02} of the $\rm{M_{bh} - M_{tot}}$ relation. In the following section, we compare our results to several black hole formation scenarios posed by various theoretical studies and examine the implications of our findings. Equation~\ref{eq:observational}, which quantifies the trend observed in Figure~\ref{fig:fromMsigma}, shows that $\rm{M_{bh}}$ scales non-linearly with $\rm{M_{tot}}$ and that efficiency of black hole formation increases with total mass. Theoretical models that reproduce the observed luminosity function of high-redshift quasars  \citep{ada03,cat01,hop05a,spr05b,vol03} predict that $\rm{M_{bh}}$ scales as a power law of the circular velocity of the galactic halo (denoted as $\rm{v_{c,halo}}$ and also referred to as virial velocity) in which the black hole resides:
\begin{eqnarray}
\rm{M_{bh}\:\propto\:{v_{c,halo}}^{\gamma}}
\label{eq:theoretical}
\end{eqnarray}

The $\rm{M_{bh} - v_{c,halo}}$ relation shown above can be converted into an equivalent $\rm{M_{bh} - M_{tot}}$ relation by considering the dependence between $\rm{v_{c,halo}}$ and halo mass (equivalent to $\rm{M_{tot}}$ in this study), $\rm{v_{c,halo}\:\propto\:{M_{halo}}^{1/3}}$. The resulting correlation between $\rm{M_{bh}}$ and $\rm{M_{halo}}$ is as follows:
\begin{eqnarray}
\rm{M_{bh}\:\propto\:{M_{halo}}^{\gamma/3}}
\label{eq:tietototalmass}
\end{eqnarray}
The circular velocity of a given halo mass is redshift-dependent; therefore, an important aspect of this analytical prediction is the evolution of the $\rm{M_{bh} - M_{halo}}$ relation with time \citep{cro09,wyi02,wyi03}. 

The slope of the $\rm{M_{bh} - M_{halo}}$ relation, $\gamma\:/\:3$, is a valuable indicator of various formation scenarios which result in observed black hole populations. In the process of hierarchical mass assembly, formation of SMBH is driven by mergers of galaxy haloes. A linear relation between $\rm{M_{bh}}$ and $\rm{M_{halo}}$, where $\gamma\:=\:3$, results from a formation scenario where the black holes residing in the merging haloes coalesce without additional gas accretion. A slope of $\gamma>3$ is characteristic of a merger where the growth of the resulting black hole is dominated by an accretion process where a significant gas fraction from the merger product is driven in to the central accreting region \citep{cat01,cat99,dim05,dim03,hae00,hae98,wyi02,wyi03}. Within this formation scenario, the fraction of baryons accreted on to the central regions that feed the black hole is significantly larger for more massive haloes, due to a deeper potential well; therefore, massive haloes host larger SMBH. 

\begin{figure}
\includegraphics[width=\textwidth]{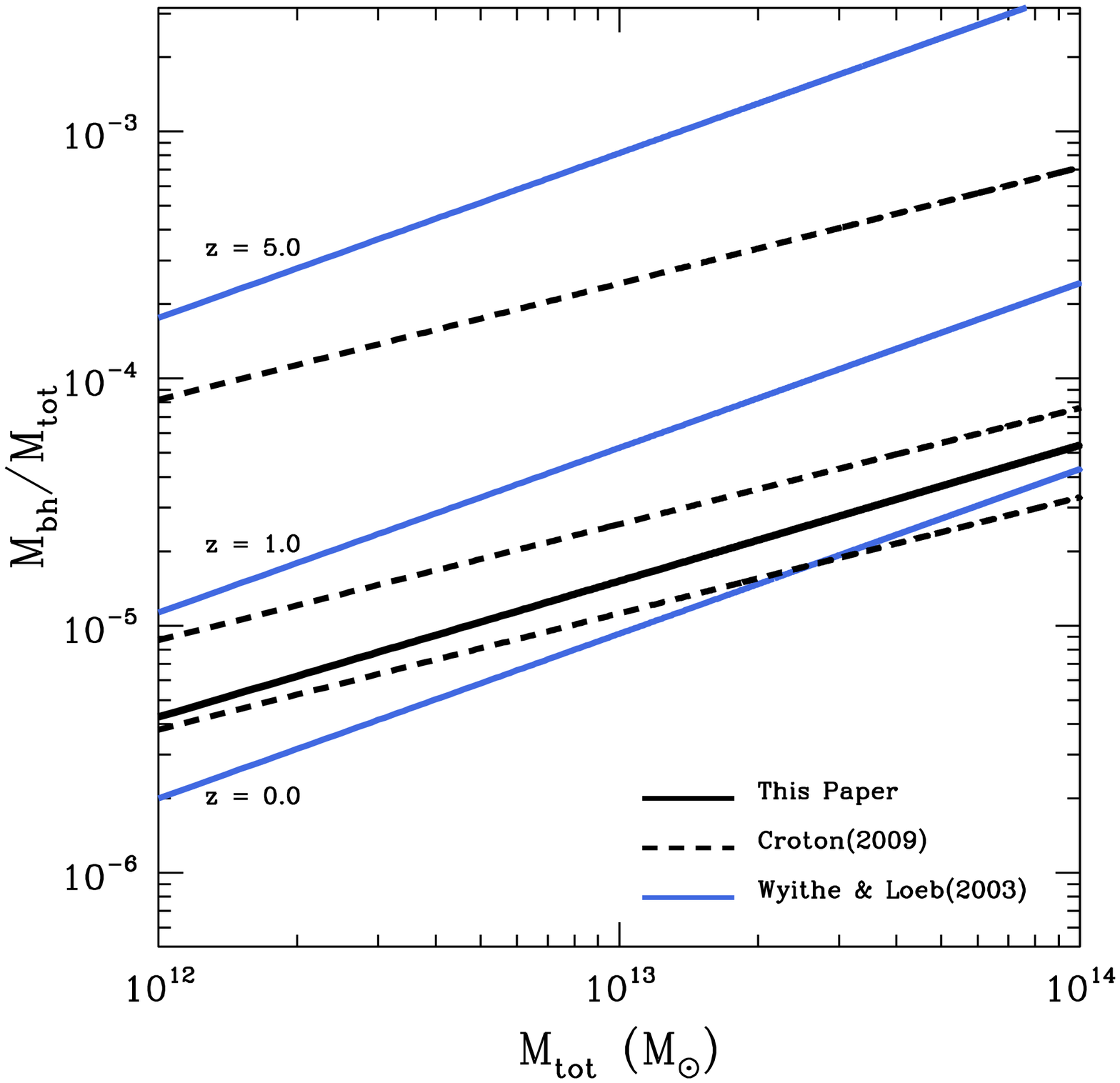}
\caption{The evolution of the $\rm{M_{bh} - M_{tot}}$ relation from various analytical predictions for $z = 0.0,1.0,5.0$. The pale blue, solid lines are predictions from \citet{wyi03}, where $\rm{M_{bh} \propto {M_{tot}}^{1.67}}$. The dashed lines are predictions from \citet{cro09}, where $\rm{M_{bh} \propto {M_{tot}}^{1.47}}$, for the same epochs as \citet{wyi03} (epochs increasing in the same order as shown in the labels for \citet{wyi03}). The thick, solid black line is a comparison of the observed $\rm{M_{bh} - M_{tot}}$ relation from the SLACS lens sample (Equation~\ref{eq:observational}), at $\langle\:z\:\rangle \approx 0.2$.}
\label{fig:combinedevolution}
\end{figure}

In Figure~\ref{fig:combinedevolution}, we plot two analytical predictions for the $\rm{M_{bh} - M_{tot}}$ relation \citep{cro09,wyi03}, which correspond to two formation scenarios, for three different epochs ($\rm{z\:=\:0.0,1.0,5.0}$). Both types of evolutionary tracks indicate that $\rm{M_{bh}}$ increases with the total mass of the host galaxy at any given redshift and that a galaxy of given total mass hosts a more massive black hole at a higher redshift relative to a lower redshift. Within the context of hierarchical mass assembly in $\rm{\Lambda}CDM$ cosmology, the decrease in the growth of $\rm{M_{bh}}$ relative to $\rm{M_{tot}}$ at lower redshifts can be caused by processes such as a decrease in merger rates and the gas fraction that is available to fuel the central SMBH. 

Another important physical process involved in the evolution of $\rm{M_{bh} - M_{tot}}$ relation is the feed-back regulated growth of SMBH \citep{dim05,hop05b,sil98,spr05a}. A black hole shines at a fraction, $\rm{\eta}$, of its Eddington luminosity ($\rm{L_{Edd}}$) following a merger and returns a fraction of this energy into the surrounding galactic gas. A black hole shining at its limiting $\rm{L_{Edd}}$ can unbind the surrounding galactic gas if the energy liberated from the black hole is sufficient to overcome the gravitational binding energy of the gas. As the mass of a black hole increases through merger driven processes discussed above, the energy output can approach the limit where it is sufficient to unbind the entire galactic gas. The unbound galactic gas that escapes into the halo is heated beyond the virial temperature and cannot cool during the dynamical time of the quasar; therefore, this mechanism can eventually terminate the accretion process that feeds the central black hole. 

The growth of a SMBH via a merger-driven, feed-back regulated mechanism implies a relation of $\rm{M_{bh}\:\propto\:{v_{c,halo}}^{5}}$ (the required rate of energy deposition to unbind a self-gravitating system is proportional to $\rm{{v_{c,halo}}^{5}/G}$) \citep{wyi02,wyi03}, leading to a slope of $\approx 1.67$ in the $\rm{M_{bh} - M_{tot}}$ relation. Evolutionary tracks shown in solid blue lines in Figure~\ref{fig:combinedevolution} \citep{wyi03} are examples of the formation scenario described above. The evolutionary tracks with a shallower slope, similar to the dashed lines shown in Figure ~\ref{fig:combinedevolution} \citep{cro09}, may be indicative of a modified feed-back regulated growth mechanism. If the galactic gas heated by the energy output of the black hole cools before the black hole reaches its critical Eddington limit, additional energy (up to a factor of $\rm{c/v_{c,halo}}$) would be required to unbind the cool gas component. This formation scenario leads to dependence of $\rm{M_{bh}\:\propto\:{v_{c,halo}}^{4}}$, resulting in a slope of $\approx 1.33$ in the $\rm{M_{bh} - M_{tot}}$ relation. It is worth noting that \citet{cro09} assumes the $\rm{M_{bh} - \sigma_{\star}}$ relation by \citet{tre02}, $\rm{M_{bh} \propto {\sigma}^{4.02}}$, which yields a relation of $\rm{M_{bh} \propto {M_{tot}}^{1.39}}$. Therefore, we recompute the results of \citet{cro09} using the $\rm{M_{bh} - \sigma_{\star}}$ relation used throughout this study \citep{gul09}, which gives a dependence of $\rm{M_{bh} \propto {M_{tot}}^{1.47}}$.

\begin{figure}
\includegraphics[width=\textwidth]{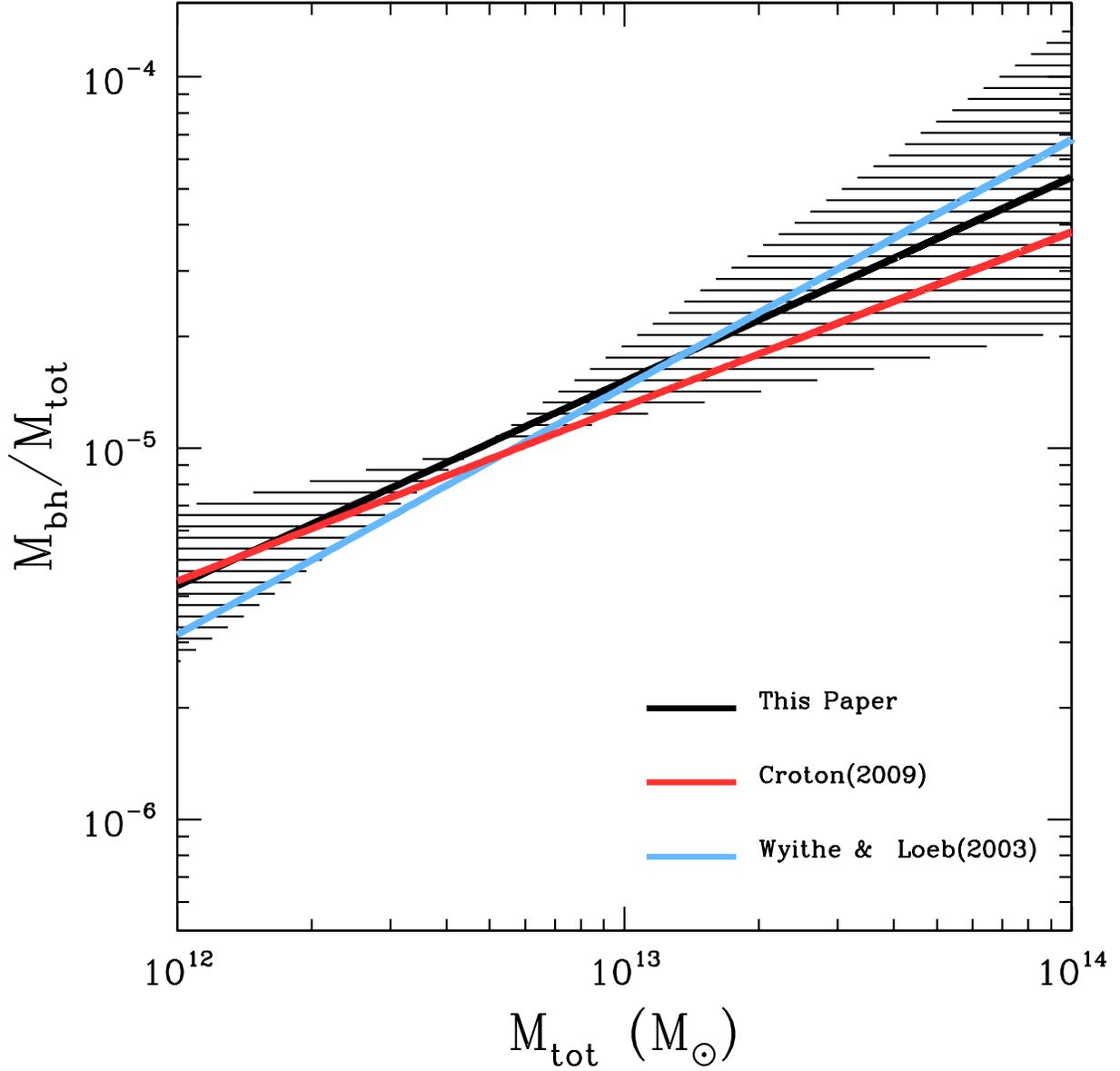}
\caption{Comparison of the observational $\rm{M_{bh} - M_{tot}}$ relation to theoretical predictions of the $\rm{M_{bh} - M_{tot}}$ relation at $\rm{z = 0.2}$, the mean redshift of the SLACS lens sample. The solid black line correspond to the best-fit result from this study and the shaded region shows the 1-$\sigma$ upper and lower limits of the $\rm{M_{bh} - M_{tot}}$ relation. The solid red and blue lines correspond to the theoretical predictions of \citet{cro09} and \citet{wyi03} of the $\rm{M_{bh} - M_{tot}}$ relation respectively.}
\label{fig:comparisonevolution}
\end{figure}

Within the context of formation and evolution of SMBH, it is extremely important to accurately determine the parameters that quantify the slope and the evolution of $\rm{M_{bh} - M_{tot}}$ relation. These parameters provide significant insight into the dominant formation scenarios that lead to the observed black hole populations at various redshifts. Due to the narrow range of redshifts in the SLACS lens sample, the $\rm{M_{bh} - M_{tot}}$ relation we derive can be considered as an evolutionary track for the mean redshift of the SLACS sample ($\langle\:z\:\rangle \approx 0.2$), as shown in Figure~\ref{fig:combinedevolution} (solid black line). The $\rm{M_{bh} - M_{tot}}$ relation we derive from the SLACS lens sample provides a unique opportunity to compare the theoretical tracks with observational evidence of the $\rm{M_{bh} - M_{tot}}$ relation. This is shown in Figure~\ref{fig:comparisonevolution}, where we compare the observational $\rm{M_{bh} - M_{tot}}$ relation to the theoretical predictions of the $\rm{M_{bh} - M_{tot}}$ relation at $\rm{z = 0.2}$. The dashed region indicates the upper and lower 1-$\rm{\sigma}$ bounds of the observational form of the $\rm{M_{bh} - M_{tot}}$ relation found in this study. Inspection of Figure~\ref{fig:comparisonevolution} shows that our results are in excellent agreement with analytical predictions of feed-back regulated growth. However, from our results, it is difficult to distinguish the importance of gas cooling in the black hole formation process.

The $\rm{M_{bh} - M_{tot}}$ relation found in this study is strongly suggestive that halo properties determine those of the galaxy and its black hole. This link is also observed in the Fundamental Plane (FP), the two-dimensional projection of the three-dimensional space defined by the quantities of  surface brightness ($\rm{I_{e}}$), effective radius ($\rm{R_{e}}$) and central velocity dispersion ($\rm{\sigma_{e}}$), of early-type galaxies \citep{djo87}. Results of \citet{bol08b} show that a sample of 53 SLACS lenses define a FP that is consistent with the general population of early-type galaxies from SDSS:
\begin{eqnarray}
\rm{R_{e}\:\propto\:{\sigma_{e}}^{1.28}\:{I_{e}}^{-0.77}}
\label{eq:slacsfp}
\end{eqnarray}
The ratio $\rm{f \equiv {\sigma_{e}} / {\sigma_{SIE}}}$  for the SLACS lenses is $\rm{f = 1.1019 \pm 0.008}$ \citep{bol08b,tre06}, indicative of a universal isothermal mass profile (also known as the ``bulge-halo conspiracy"). Therefore, we can replace the quantities $\rm{R_{e}}$ and $\rm{\sigma_{e}}$ in the FP with $\rm{R_{200}}$ (given the trend observed between the quantities $\rm{log(M_{einst} / M_{200})}$ and $\rm{log(R_{einst} / R_{200})}$) and $\rm{\sigma_{SIE}}$. Combining this with equations~\ref{eq:totalmass} and ~\ref{eq:Msigma}, we derive:
\begin{eqnarray}
\rm{M_{bh} \propto {M_{tot}}^{1.30}\:{I_{e}}}
\label{eq:slacsFP2}
\end{eqnarray}
Ignoring the dependence of equation~\ref{eq:slacsFP2} on $\rm{I_{e}}$, since it only varies weakly relative to the other variables, we can extract the connection between $\rm{M_{bh}}$ and $\rm{M_{tot}}$ from the FP of SLACS lenses. The slope of 1.30 of the $\rm{M_{bh} - M_{tot}}$ relation derived from the SLACS FP \citep{bol08b} is in agreement with our results within the 1-$\rm{\sigma}$ bounds. The manifestation of a $\rm{M_{bh} - M_{tot}}$ relation within the FP of the SLACS lenses further strengthens the existence of the observed $\rm{M_{bh} - M_{tot}}$ relation. We require $\it{direct}$ measurements of both $\rm{M_{bh}}$ and  $\rm{M_{tot}}$ to fully compare the manifestation of $\rm{M_{bh} - M_{tot}}$ relation within the FP and its implications for regularity of early-type galaxy formation scenarios.

An additional advantage of the results from this study lies within the method used to derive $\rm{M_{tot}}$. The traditional method of estimating total gravitational mass of a galaxy, used in black hole studies, is to convert the ${\it observed}$ circular velocity (from galaxy rotation curves) into the velocity of the galactic halo (also referred to as the ${\it virial}$ velocity). The inferred total mass of a galaxy differs depending on the method used to relate observed circular velocity to the virial velocity. The effect on the $\rm{M_{bh} - M_{tot}}$ relation from varying assumptions of the connection between observed circular velocity ($\rm{v_{c,obs}}$) and virial velocity ($\rm{v_{vir}}$), discussed in \citet{fer02}, is also shown in Figure~\ref{fig:fromMsigma}. The dashed line shows the best-fit $\rm{M_{bh} - M_{tot}}$ relation obtained by \citet{fer02} using cosmological prescriptions of \citet{bul01} to relate $\rm{v_{c,obs}}$ and $\rm{v_{vir}}$ ($\rm{M_{bh} \propto {M_{tot}}^{1.65}}$) and the dot-dashed line shows the resulting relation where $\rm{v_{c,obs}\:=\:1.8\:v_{vir}}$ ($\rm{M_{bh} \propto {M_{tot}}^{1.82}}$). From the relations obtained by \citet{fer02}, it is evident that the slope of the $\rm{M_{bh} - M_{tot}}$ is affected by the method assumed to relate $\rm{v_{c,obs}}$ and $\rm{v_{vir}}$. Gravitational lens modeling is independent of such dynamical assumptions and provides an elegant alternative method to determine the total mass. Therefore, the $\rm{M_{bh} - M_{tot}}$ relation characterized by this independent method is not only a complementary comparison to existing observational evidence but also provides valuable insight to determine the dominant physical processes of SMBH growth. 

\section{CONCLUSIONS}

We use a sample of 43 early-type galaxies, which exhibit galaxy-scale strong gravitational lensing, to derive the scaling relation between black hole mass, $\rm{M_{bh}}$, and the total mass, $\rm{M_{tot}}$, of the host galaxy. In this study, we use gravitational lens modeling to directly measure the total mass and the mass profile of the galaxy rather than converting the observed circular velocity into a total mass, the traditional method that is used in black hole studies. We use two alternative scaling relations, $\rm{M_{bh} - \sigma_{\star}}$ and $\rm{M_{bh}} - {\it n}$ to estimate the black hole masses of the lens sample. We obtain a tight correlation between $\rm{\sigma_{\star}}$ and $\rm{M_{tot}}$ in the log-log space. In conjunction with the $\rm{M_{bh} - \sigma_{\star}}$ relation, we derive the observational form of the $\rm{M_{bh} - M_{tot}}$ relation that is consistent with no intrinsic scatter. We do not find a significant correlation between $n$ and $\rm{M_{tot}}$. From a variety of tests, we find that we cannot confidently estimate black hole masses with the $\rm{M_{bh}} - n$ relation. The scaling relation we observe between $\rm{M_{bh}}$ and $\rm{M_{tot}}$ is non-linear and is in agreement with theoretical predictions of the growth of black holes and observational studies of the local $\rm{M_{bh} - M_{tot}}$ relation. The observed $\rm{M_{bh} - M_{tot}}$ relation is also consistent with the studies of the Fundamental Plane of SLACS lenses \citep{bol08b}, which is suggestive of a unified scenario where the properties of the host halo determine those of the resulting galaxy and black hole formed through hierarchical merging. The observed non-linear correlation between $\rm{M_{bh}}$ and $\rm{M_{tot}}$ indicates that massive halos are more efficient in forming black holes and the slope of the $\rm{M_{bh} - M_{tot}}$ relation is suggestive of a merger-driven, feed-back regulated process for the growth of black holes.

\acknowledgments
 
 We wish to thank Laura Ferrarese, Chien Peng, Christopher Pritchet, Jon Willis, and the members of the SLACS team for valuable discussions and useful comments. We also wish to thank the anonymous referee whose helpful comments improved this paper. This research used the facilities of the Canadian Astronomy Data Centre operated by the National Research Council of Canada with the support of the Canadian Space Agency. The authors gratefully acknowledge the support from the National Research Council of Canada and NSERC through Discovery grants.

\clearpage
 
\clearpage
 
\clearpage

\clearpage
 
\clearpage

\clearpage

\clearpage

\clearpage

\clearpage

\clearpage

\clearpage

\begin{deluxetable}{lccccccc}
\tabletypesize{\scriptsize}
\tablewidth{0pt}
\setlength{\tabcolsep}{0.1in}
\tablenum{1}
\tablecaption{Galaxy Sample}
\tablehead{ \colhead{System} & \colhead{$z_{lens}$} & \colhead{$b_{frac}$} & \colhead{S$\rm{\acute{e}}$rsic (${\it n}$)} & \colhead{$\rm{M_{bh,n}}$ ($\rm{M_{\odot}}$)} & \colhead{$\sigma_{\rm{ap}}\:(\rm{km/s})$} & \colhead{$\rm{M_{bh,\sigma}}$ ($\rm{M_{\odot}}$)} & \colhead{$\rm{M_{200}}$ ($\rm{M_{\odot}}$)}\\
 \colhead{(1)} & \colhead{(2)} & \colhead{(3)} & \colhead{(4)} & \colhead{(5)} & \colhead{(6)} & \colhead{(7)} & \colhead{(8)}}
\startdata
SDSS J0029-0055 & 0.23 & 0.76 & $6.41 \pm 1.28$ & 7.29\e{8} & $235.86 \pm 18.54$ & 2.65\e{8} & 1.04\e{13} \\
SDSS J0037-0942 & 0.20 & 0.94 & $4.47 \pm 0.89$ & 3.38\e{8} & $282.43 \pm 14.17$ & 5.69\e{8} & 2.40\e{13} \\
SDSS J0216-0812 & 0.33 & 0.43 & $2.91 \pm 0.58$ & 8.50\e{7} & $356.24 \pm 24.61$ & 1.52\e{9} & 4.03\e{13} \\
SDSS J0252+0039 & 0.28 & 0.58 & $10.64 \pm 2.12$ & 1.19\e{9} & $169.18 \pm 12.38$ & 6.48\e{7} & 1.27\e{13} \\
SDSS J0330-0020 & 0.35 & 0.88 & $7.85 \pm 1.85$ & 9.65\e{8} & $217.42 \pm 21.54$ & 1.88\e{8} & 1.51\e{13} \\
SDSS J0728+3835 & 0.21 & 0.95 & $6.36 \pm 1.27$ & 7.21\e{8} & $218.03 \pm 11.21$ & 1.90\e{8} & 1.73\e{13} \\
SDSS J0737+3216 & 0.32 & 0.85 & $6.10 \pm 1.22$ & 6.70\e{8} & $341.91 \pm 17.20$ & 1.28\e{9} & 2.41\e{13} \\
SDSS J0822+2652 & 0.24 & 0.81 & $4.98 \pm 1.00$ & 4.41\e{8} & $269.55 \pm 15.61$ & 4.67\e{8} & 2.00\e{13} \\
SDSS J0912+0029 & 0.16 & 1.00 & $3.30 \pm 0.66$ & 1.35\e{8} & $329.39 \pm 16.17$ & 1.09\e{9} & 4.36\e{13} \\
SDSS J0935-0003 & 0.35 & 0.27 & $2.75 \pm 0.55$ & 6.89\e{7} & $428.32 \pm 37.36$ & 3.33\e{9} & 4.46\e{13} \\
SDSS J0936+0913 & 0.19 & 0.79 & $5.88 \pm 1.18$ & 6.27\e{8} & $250.43 \pm 12.37$ & 3.42\e{8} & 1.48\e{13} \\
SDSS J0946+1006 & 0.22 & 0.40 & $2.19 \pm 0.44$ & 2.61\e{7} & $284.95 \pm 22.75$ & 5.91\e{8} & 2.32\e{13} \\
SDSS J0955+0101 & 0.11 & 0.77 & $8.57 \pm 1.71$ & 1.05\e{9} & $193.13 \pm 13.08$ & 1.14\e{8} & 1.21\e{13} \\
SDSS J0956+5100 & 0.24 & 0.90 & $7.72 \pm 1.54$ & 9.48\e{8} & $330.61 \pm 16.83$ & 1.11\e{9} & 3.24\e{13} \\
SDSS J0959+0410 & 0.13 & 0.14 & $2.09 \pm 0.42$ & 2.10\e{7} & $226.97 \pm 14.98$ & 2.25\e{8} & 1.08\e{13} \\ 
SDSS J0959+4416 & 0.24 & 0.64 & $3.65 \pm 0.73$ & 1.87\e{8} & $259.69 \pm 20.22$ & 3.99\e{8} & 1.65\e{13} \\
SDSS J1020+1122 & 0.28 & 0.58 & $4.38 \pm 0.88$ & 3.20\e{8} & $298.15 \pm 19.03$ & 7.16\e{8} & 2.76\e{13} \\
SDSS J1029+0420 & 0.10 & 0.84 & $5.87 \pm 1.17$ & 6.24\e{8} & $215.58 \pm 11.29$ & 1.81\e{8} & 9.82\e{12} \\
SDSS J1106+5228 & 0.10 & 0.79 & $5.70 \pm 1.14$ & 5.90\e{8} & $273.07 \pm 13.55$ & 4.94\e{8} & 1.49\e{13} \\
SDSS J1112+0826 & 0.27 & 0.67 & $2.62 \pm 0.52$ & 5.63\e{7} & $344.43 \pm 21.53$ & 1.32\e{9} & 3.08\e{13} \\
SDSS J1134+6027 & 0.15 & 0.53 & $3.44 \pm 0.69$ & 1.55\e{8} & $256.76 \pm 12.89$ & 3.80\e{8} & 1.51\e{13} \\
SDSS J1142+1001 & 0.22 & 0.48 & $3.71 \pm 0.74$ & 1.98\e{8} & $238.99 \pm 23.79$ & 2.81\e{8} & 1.68\e{13}\\
SDSS J1143-0144 & 0.11 & 0.90 & $3.38 \pm 0.68$ & 1.46\e{8} & $269.91 \pm 13.04$ & 4.70\e{8} & 2.52\e{13} \\
SDSS J1204+0358 & 0.16 & 0.72 & $4.96 \pm 0.99$ & 4.37\e{8} & $283.45 \pm 18.05$ & 5.78\e{8} & 1.72\e{13} \\
SDSS J1205+4910 & 0.22 & 0.78 & $5.58 \pm 1.12$ & 5.66\e{8} & $287.77 \pm 14.34$ & 6.17\e{8} & 2.37\e{13} \\
SDSS J1213+6708 & 0.12 & 0.89 & $10.41 \pm 2.08$ & 1.19\e{9} & $283.14 \pm 14.54$ & 5.76\e{8} & 1.70\e{13} \\
SDSS J1250+0523 & 0.23 & 0.70 & $8.57 \pm 1.71$ & 1.05\e{9} & $258.15 \pm 14.34$ & 3.89\e{8} & 1.46\e{13} \\
SDSS J1402+6321 & 0.20 &1.00 & $5.94 \pm 1.19$ & 6.38\e{8} & $265.73 \pm 16.92$ & 4.40\e{8} & 2.61\e{13} \\
SDSS J1403+0006 & 0.19 & 0.66 & $10.23 \pm 2.05$ & 1.18\e{9} & $213.50 \pm 17.04$ & 1.74\e{8} & 1.18\e{13} \\
SDSS J1416+5136 & 0.30 & 0.55 & $2.67 \pm 0.53$ & 6.07\e{7} & $260.54 \pm 27.14$ & 4.04\e{8} & 2.31\e{13} \\
SDSS J1420+6019 & 0.06 & 0.75 & $3.27 \pm 0.65$& 1.29\e{8} & $213.46 \pm 10.41$ & 1.74\e{8} & 9.39\e{12} \\
SDSS J1430+4105 & 0.29 & 0.74 & $4.28 \pm 0.86$ & 3.00\e{8} & $335.24 \pm 33.32$ & 1.18\e{9} & 3.76\e{13} \\
SDSS J1443+0304 & 0.13 & 0.89 & $10.04 \pm 2.01$ & 1.17\e{9} & $213.43 \pm 11.23$ & 1.74\e{8} & 9.46\e{12} \\
SDSS J1451-0239 & 0.13 & 0.93 & $5.49 \pm 1.10$ & 5.46\e{8} & $225.94 \pm 14.18$ & 2.21\e{8} & 1.17\e{13} \\
SDSS J1525+3327 & 0.36 & 0.58 & $4.40 \pm 0.88$ & 3.23\e{8} & $274.71 \pm 27.05$ & 5.06\e{8} & 3.03\e{13} \\
SDSS J1531-0105 & 0.16 & 1.00 & $5.45 \pm 1.09$ & 5.37\e{8} & $280.78 \pm 14.09$ & 5.55\e{8} & 2.35\e{13} \\
SDSS J1538+5817 & 0.14 & 0.86 & $9.31 \pm 1.86$ & 1.12\e{9} & $189.86 \pm 12.05$ & 1.06\e{8} & 1.17\e{13} \\
SDSS J1627-0053 & 0.21 & 0.67 & $4.42 \pm 0.88$ & 3.27\e{8} & $305.19 \pm 15.79$ & 7.91\e{8} & 2.11\e{13} \\
SDSS J1630+4520 & 0.25 & 0.86 & $4.11 \pm 0.82$ & 2.68\e{8} & $287.64 \pm 16.67$ & 6.15\e{8} & 3.02\e{13} \\
SDSS J1636+4707 & 0.23 & 0.64 & $3.29 \pm 0.66$ & 1.33\e{8} & $247.07 \pm 16.04$ & 3.23\e{8} & 1.53\e{13} \\
SDSS J2238-0754 & 0.14 & 0.80 & $4.49 \pm 0.88$ & 3.41\e{8} & $205.23 \pm 11.40$ & 1.47\e{8} & 1.44\e{13} \\
SDSS J2303+1422 & 0.16 & 0.97 & $4.13 \pm 0.83$ & 2.71\e{8} & $257.14 \pm 16.13$ & 3.83\e{8} & 2.57\e{13} \\
SDSS J2341+0000 & 0.19 & 0.52 & $7.43 \pm 1.49$ & 9.02\e{8} & $204.24 \pm 12.83$ & 1.44\e{8} & 1.87\e{13}
\enddata
\tablecomments{Col. (1): The unique SDSS spectrum identifier for the lens system. Col (2): Redshift of the early-type lens galaxy \citep{bol08} Col. (3): Bulge fraction of the early-type galaxy from the best-fit bulge-disk decomposition. Col (4): S$\rm{\acute{e}}$rsic index from the best-fit bulge-disk galaxy decomposition. Col. (5): black hole mass estimated from the $\rm{M_{bh}} -{\it n}$ relation. The 1-$\sigma$ error on the quantity $\rm{log\:(M_{bh,n})}$ is a function of $n$ and its measurement errors and can be computed as follows \citep{gra07}: $\rm{\delta\:log(M_{bh}) = \sqrt{[log(n/3)]^{4} + \frac{[log(n/3)]^{2}}{4} + 0.09^{2} + \frac{[3.70 - 6.20\:log(n/3)]^{2}\:(\delta n/n)^{2}}{(ln 10)^{2}} + 0.18^{2}}}$ Col. (6): Velocity dispersion obtained from the SDSS database corrected to a standard aperture of $\rm{r_{e}/8}$. Col. (7): black hole mass estimated from the $\rm{M_{bh} - \sigma}$ relation. The 1-$\sigma$ error on the quantity $\rm{log\:(M_{bh,\sigma_{\star}})}$ is a function of $\sigma_{\star}$ and its measurement errors and can computed as follows: $\rm{\delta(log(M_{bh})) = \sqrt{0.08^{2} + (log(\sigma_{\star}/ 200\:km\:s^{-1}))^{2}\:(0.41^{2}) + (\delta\:log(\sigma_{\star}/ 200\:km\:s^{-1}))^{2}\:(4.24^{2}) + 0.44^2}}$ Col. (8): Total mass of the lensing galaxy within the projected radius of $\rm{R_{200}}$, where mass profile of the lensing galaxy is derived from the lens modeling results of \citet{bol08}. The typical 1-$\sigma$ error on the quantity $\rm{log\:(M_{tot})}$ is $\approx 0.01\:\rm{dex}$.}
\end{deluxetable}

\clearpage

\end{document}